\documentclass{osa-article}
\journal{osac}
\articletype{Research Article}

\begin{document}

\title{Terahertz waves generation using the isomorphs of PPKTP crystal: a theoretical investigation}

\author{Guo-Qun Chen,\authormark{1} Hong-Yang Zhao,\authormark{1} Shun Wang,\authormark{1} Xiangying Hao,\authormark{1,*} Hai-Wei Du,\authormark{2} and Rui-Bo Jin\authormark{1,\dag}}

\address{\authormark{1}Hubei Key Laboratory of Optical Information and  Pattern Recognition, and Department of Materials Science and Engineering, Wuhan Institute of Technology, Wuhan 430205, China\\
\authormark{2}School of Measuring and Optical Engineering, Nanchang Hangkong University, Nanchang, Jiangxi 330063, China\\}

\email{\authormark{*}xyhao.321@163.com} 
\email{\authormark{\dag}jrbqyj@gmail.com}



\begin{abstract}
Highly efficient terahertz (THz) wave sources based on difference frequency generation (DFG) process in
nonlinear optical crystals play an important role for the applications of  THz wave. In order to find more
novel nonlinear crystals, here we theoretically investigate the generation of THz wave using the isomorphs
of periodically poled $\mathrm{KTiOPO_4}$ (PPKTP), including periodically poled RTP, KTA, RTA and CTA. By
solving the cascaded difference frequency coupled wave equations, it is found that the intensities of
the THz wave generated from the cascaded difference frequency processes are improved by 5.27, 2.87, 2.82,
3.03, and 2.76 times from the non-cascaded cases for KTP, RTP, KTA, RTA and CTA, respectively. The effects
of the crystal absorption,  the phase mismatch  and the pump intensity are also analyzed in detail. This
study might help to provide a stronger THz radiation source based on the nonlinear crystals.
\end{abstract}
\section{Introduction}
High-power and high-stability terahertz (THz) wave sources have received extensive attention due to their
important applications in time-domain spectroscopy system, imaging, material detection, biomedical
science, etc \cite{Zhang2010book, Lee2009, Zhong2011, Kawase2007, Murate2018}. There are several methods
to generate THz radiation, for example, optical parametric oscillation \cite{Li2013}, photoconductive
antenna \cite{Burford2017}, optical rectification \cite{Du2015} and optical difference frequency
generation (DFG) \cite{Ding2014}. Among these methods, optical DFG is widely used due to its advantages of
wide tuning range, high output power, and running at room temperature, etc. In an optical DFG process,
two tunable near-infrared lasers with similar frequencies  are combined and interact in a nonlinear
crystal. However, the traditional DFG method is limited by the order of difference frequency, and the
conversion efficiency of THz wave is very low, even lower than the absorption effect of the crystal. To
solve this problem, the method of cascaded DFG  was proposed \cite{Liu2013} and this method has the merits
of higher quantum conversion efficiency and overcoming the limitation of Manley-Rowe condition
\cite{Saito2015}. Manley-Rowe condition means the highest achievable conversion efficiency is the quotient
of output and input frequencies in the conventional optical conversion \cite{Waldmueller2007}.

The periodically poled KTP crystal (PPKTP) is one of the most widely used crystals in nonlinear optics
because it has wide transparency in the range of visible light and THz wave, and a relatively high
nonlinear coefficient (15.4 pm/V at 1064 nm) and a high light damage threshold (10 J/cm$^2$ at 1064
nm)\cite{Hildenbrand2009}. PPKTP has many isomorphic crystals, such as RTP, KTA, RTA and CTA, which have
similar properties as  their "parent" KTP. PPKTP and its isomorphs have been applied for extended phase
matching \cite{Kim2015JKPS, Kim2021}, the generation of spectrally pure single photons sources at telecom
wavelength \cite{Jin2016PRAppl, Laudenbach2017, Jin2019OLT} and  for mid-infrared single photon generation
\cite{McCracken2018}.

Recently, the generation THz wave using PPKTP and PPRTP in cascade DFG process has been theoretically
investigated and  achieved the THz intensity enhancement of 5.53 and 2.95 times, respectively
\cite{Li2017, Li2017a}. These studies provide new possibilities in the direction of searching for new
crystals with higher nonlinear coefficient for higher quantum efficiency. Inspired by the previous works \cite{Liu2013JLT, Wang2015MRLB, Lu2018OE, Tian2021}, here we study the THz generation from the other three isomorphs of KTP, namely KTA, RTA and CTA.
We compare these three isomorphs with KTP and RTP in Table \ref{table1}. Some  optical properties of these
three crystals are even better than KTP and RTP, especially the effective nonlinear coefficient of KTA,
RTA and CTA are larger than KTP and RTP. So, we theoretically expect these crystals  can also achieve high
quantum conversion efficiency for generating THz wave. These isomorphic crystals may provide more options
for the preparation of terahertz wave in experiment in the future.

This article consists of the following parts: Section 1 is the introduction part; In Section 2, we
establish the theoretical model of cascaded DFG in the form of  coupled wave equations. In Section 3, we
first calculate the wave vector mismatch and THz intensity under cascade conditions, and then we compare
the effects of different conditions on the THz intensity and the quantum conversion efficiency. We provide
comprehensive discussions in Section 4, and finally summarize the article in Section 5.
\begin{table*}[tbp]
\centering
\begin{tabular}{cccccc}
\hline
\hline
Name&PPKTP & PPRTP & PPKTA & PPRTA & PPCTA \\

Composition& $KTiOPO_4$ & $RbTiOPO_4$ & $KTiOAsO_4$ & $RbTiOAsO_4$ & $CsTiOAsO_4$ \\
\hline
$\alpha_n$($cm^{-1}$) & 0.0005 & 0.05 & 0.005 & $0.005$ & $0.005$ \\
$\alpha_T$($cm^{-1}$) & 1.69 & 4.04 & 3.79 & 3.59 & 4.31 \\
$d_{eff}$(pm/V)& 15.4 & 17.1 & 16.2 & 17.4 & 18.1 \\
$n_p$@282 THz& 1.8297 & 1.85302 & 1.86764 & 1.88087 & 1.91924 \\
$n_s$@281 THz& 1.82956 & 1.85286 & 1.86749 & 1.88072 & 1.91908 \\
$n_T$@1 THz& 3.80761 & 3.34908 & 4.02779 & 4.29029 & 3.97803 \\
$\Lambda$($\mu$m)& 236.854 & 219.902 & 236.129 & 241.549 & 225.387 \\
References&\cite{Perkins1987, Gettemy1988, Kugel1988, Shoji1997, Kato2002, Sang2010}&\cite{Sang2010, Cheng1994, Hansson2000, Mikami2009}&\cite{Hansson2000, Watson1991, Loiacono1992, Cheng1993a, Kato1997,  Mounaix2004}&\cite{Hansson2000, Cheng1993b, Guo1996, Yang1999,  Kato2003}&\cite{Guo1996, Cheng1993, Loiacono1993,  Mikami2011} \\
\hline
\hline
\end{tabular}
\caption{ Main parameters of PPKTP and its isomorphic crystals. $\alpha_n$ and $\alpha_T$ are absorption coefficients in optical band and THz band, respectively.  $d_{eff}$ is effective nonlinear coefficient. $n_p$, $n_s$ and $n_T$ are the refractive indices for the pump, signal and THz wave. $\Lambda$ is the poling period.}
\label{table1}
\end{table*}

\section{Theory}
\subsection{Characteristics of PPKTP and its isomorphic crystals}
The five crystals that we use to generate THz radiation from the PPKTP family are listed in Table
\ref{table1}. They have the same point group (mm2) in structure, similar lattice constants
(a$\approx$1.281 nm, b$\approx$0.6404 nm, c$\approx$1.0616 nm) and similar transparency (in the range of
0.35-5.3$\mu$m). They are isomorphic to each other, so they are close in characteristics. The isomorphs of
PPKTP crystal have higher light damage thresholds than conventional nonlinear optical crystals (e.g. PPLN,
ZnTe, and GaAs) \cite{Du2014, Cronin-Golomb2004, Schaar2008}. In addition, PPKTP and its isomorphic
crystals have lower refractive indexes  in the THz band, which is conducive to the generation of
high-power THz waves in the DFG processes.

In order to simulate the performance of these five crystals in THz generation, we summarize the key
parameters in Table \ref{table1}, where the initial frequency of pump and signal laser are 282 and 281
THz, corresponding to wavelength of about 1064 nm and 1067 nm respectively, THz wave frequency is 1 THz.
Note, the absorption coefficients $\alpha_n$ for KTP, RTP and KTA are obtained from previous experimental
data, while the $\alpha_n$  for RTA and CTA in optical band are still not reported in previous
literatures. So, we theoretically assumed that RTA and CTA  have the same  value as  KTA. The RTA and CTA are chosen to have closer absorption coefficients as KTA instead of RTP and KTP, because RTA, CTA, and KTA have the same chemical unit of $(TiOAsO_4)^{-1}$. The refractive indices of the pump, the signal, and the THz wave in the crystal are obtained by using the Sellmeier equations respectively.

\begin{figure}[tbh]
\centering\includegraphics[width=10cm]{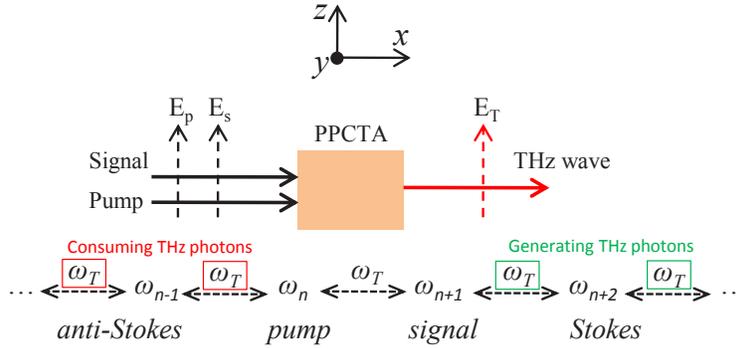}
\caption{Schematic diagram of cascaded difference frequency. The propagation directions of the pump, the signal and THz wave are along the x axis, while their polarization directions are along the z axis. }
\label{figure1}
\end{figure}

\subsection{Theory of cascaded DFG}
We assume the frequencies of the incident pump light, the signal light, and the THz wave are $\omega_n$,
$\omega_{n+1}$ and $\omega_T$, respectively. Under the nonlinear interactions in a cascaded DFG, a series
of waves with frequencies of ..., $\omega_{n-1}$, $\omega_n$, $\omega_{n+1}$, $\omega_{n+2}$, ... are
generated, and their mutual interval is $\omega_T$. The light with frequency higher than the incident
frequency ($\omega_{n+1}$, $\omega_{n+2}$...) are called anti-Stokes light, and the light with  frequency
(...$\omega_{n-2}$, $\omega_{n-1}$) lower than the incident frequency called Stokes light, as shown
in Fig.\ref{figure1}.

The coupled wave equations of cascaded DFG can be derived from the nonlinear parametric interaction
coupled wave equations\cite{Ravi2016}, with a specific form as follow\cite{Li2017}:
\begin{equation}\label{eq1}
\frac{dE_T}{dz}=-\frac{\alpha_T}{2}E_T+\kappa_T\sum_{n=-\infty}^{+\infty}E_nE_{n+1}\cos{(\Delta k_nz)},
\end{equation}
and
\begin{equation}\label{eq2}
\frac{dE_n}{dz}= -\frac{\alpha_n}{2}E_n+\kappa_nE_{n-1}E_T\cos{(\Delta k_{n-1}z)}
-\kappa_nE_{n+1}E_T\cos{(\Delta k_nz)},
\end{equation}
where $E_n$ and $E_T$ respectively represent the electric fields of the pump light and the THz wave. The
second and third terms on the right hand side of Eq. (\ref{eq2}) represent the Stokes process (generating
THz photons) and the anti-Stokes process (consuming THz photons), respectively. In our theory, the
direction of the electric field of the pump light, signal light and THz wave are the same (see Fig.
\ref{figure1}), $\alpha_n$ and $\alpha_T$ are the absorption coefficients of the pump light and the THz
wave in the crystal respectively, $\Delta k_n$ represents the phase mismatch in the cascade process,
$\kappa_n$ and $\kappa_T$ are the coupling coefficients. The phase mismatch and the coupling coefficient
can be calculated using the following equations \cite{Li2017}:

\begin{equation}\label{eq3}
\Delta k_n=k_n-k_{n+1}-k_T-\frac{2\pi}{\Lambda},
\end{equation}
\begin{equation}\label{eq4}
\kappa_n=\frac{\omega_nd_{eff}}{cn_\mu},
\end{equation}
\begin{equation}\label{eq5}
\kappa_T=\frac{\omega_Td_{eff}}{cn_T},
\end{equation}
where $d_{eff}$ is the effective nonlinear coefficient, $n_\mu$ ($\mu$ = p, s) and $n_T$ are the
refractive indices of pump (signal) light and THz wave, respectively, and $\Lambda$ is the poling period of the nonlinear crystal. Combined with the equations (1) and (2), the THz power density in the cascaded difference frequency processes can be calculated.

\section{Numerical simulations}
First, we consider the parameter of coherence length, which is defined as the distance of the second harmonic intensity reaching its maximum for the first time.
Coherence length is an important parameter for optical pulses.
The coherence length $\emph L_c$ and the phase mismatch $\Delta k$ in PPKTP and its isomorphic crystals
can be calculated according to Eq. (\ref{eq3}). The relationship between $\emph L_c$ and  $\Delta k$  can
be described by the following equation\cite{Li2017a}:
\begin{equation}\label{eq10}
L_c=\frac{\pi}{\Delta k}.
\end{equation}
\begin{figure*}[htbp]
\centering\includegraphics[width=13cm]{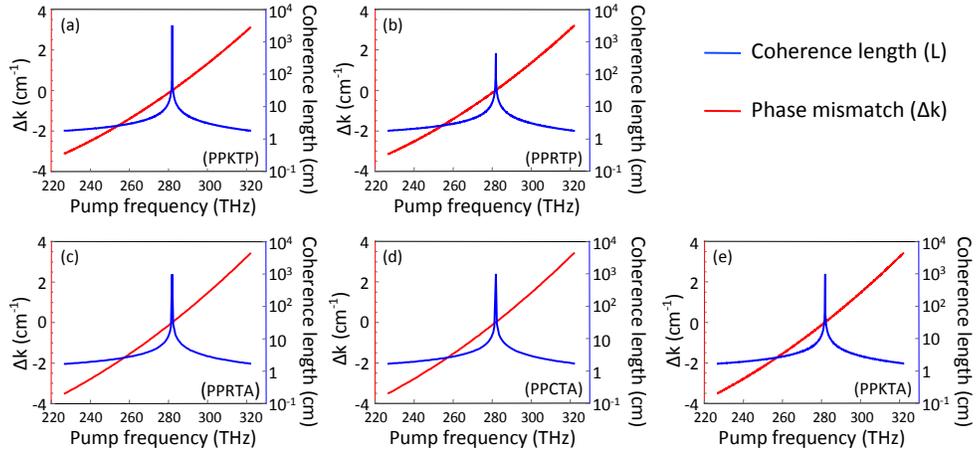}
\caption{Phase mismatch and coherence length of cascaded difference frequency processes(Blue line: coherence length; Red line: phase mismatch).
 } \label{figure2}
\end{figure*}
The simulation for $\emph L_c$ and  $\Delta k$  is shown in Fig.\ref{figure2}.
In the simulation, the frequencies of the pump light and the signal light are set to be 282 and 281 THz,
respectively; the THz wave frequency is 1 THz and the range of frequency is from 227 THz to 322 THz.  See Mathematica codes in Supplementary Information.

In Fig.\ref{figure2}, the value of $\Delta k$ approaches 0 near the initial frequency; the part lower
than the initial frequency ($\Delta k < 0$, 227-280 THz) represents the Stokes process, and the part
higher than the initial frequency ($\Delta k > 0$, 283-322 THz) represents the anti-Stokes process. As the
cascade process continues, the gap between the pump wave vector ($k_n$) and the signal wave vector
($k_{n+1}$) increases, and the absolute value of $\Delta k$ also increases gradually. Simultaneously, the
coherence length reaches the maximum near the initial frequency, and then decreases rapidly. Here,  $\emph
L_c$ is shown in an exponent with a base of 10.
\begin{figure*}[htbp]
\centering\includegraphics[width=13cm]{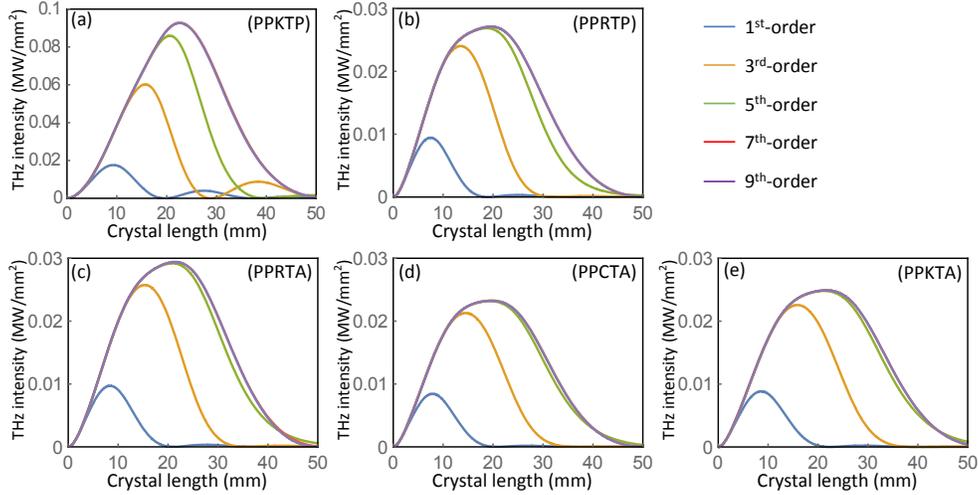}
\caption{THz wave intensity in non-cascaded and cascaded processes (initial frequency $\omega_p$=282THz,
$\omega_s$=281THz).
 } \label{figure3}
\end{figure*}

In order to observe the influence of cascaded DFG on THz wave intensity, we compare the THz wave intensity
generated by PPKTP, PPRTP, PPKTA, PPRTA, and PPCTA in the case of non-cascaded DFG and cascaded DFG in
Fig.\ref{figure3} ($1^{\rm{st}}$, $3^{\rm{rd}}$, $5^{\rm{th}}$, $7^{\rm{th}}$, and $9^{\rm{th}}$ order),
respectively. It can be seen that the intensity of the THz wave in the case of no cascade (i.e. the
$1^{\rm{st}}$ order line) is very low. And it can be calculated that the THz intensity at $9^{\rm{th}}$-order cascaded DFG is 5.27, 2.82, 2.87, 3.03 and 2.76 times of the case of non-cascaded DFG,
respectively. The results indicate that utilizing the isomorphs of PPKTP crystal and cascaded DFG method
is feasible for increasing the intensity of the THz wave.

In Fig.\ref{figure3}, the curves of the  $7^{\rm{th}}$-order cascade and the $9^{\rm{th}}$-order
cascade are almost overlapped, indicating that the THz wave intensity at the $9^{\rm{th}}$-order reaches
the maximum.
In this calculation, we give a same initial value 1 MW/mm$^2$ of the pump and the signal intensity to calculate the THz intensity.
 In addition, the maximal THz intensity in Fig. \ref{figure3}(b, c, d, e) are comparable,
i.e., between 0.02 and 0.03 MW/mm$^2$, lower than the value in Fig. \ref{figure3}(a), which is mainly due
to the absorption coefficient of PPKTP crystal in THz band is smaller than other crystals.
Another interesting phenomenon is that a second peak appears when the cascade process continues in the
PPKTP crystal (the phenomenon also exists in four crystals). This is because the THz wave is converted to
pump light (sum frequency) when its intensity reaches the first peak in the cascade process, and then the
difference frequency process continues to generate THz wave, but owing to the crystal absorption, the
second peak is much weaker than the first one.

\begin{figure*}[tbh]
\centering\includegraphics[width=12cm]{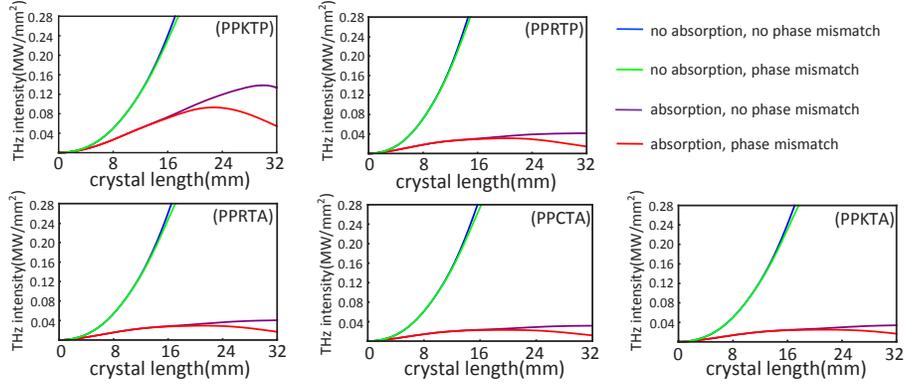}
\caption{THz intensity versus crystal length with different crystal absorption and phase mismatch (Blue line: with no absorption and no phase mismatch; Green line: with no absorption but phase mismatch;  Purple line: with absorption  but no phase mismatch; Red line: with absorption and mismatch).
 } \label{figure4}
\end{figure*}

In the process of cascaded DFG, the crystal absorption and the phase mismatch are two most important factors affecting the THz intensity, so next we compare the variation of THz intensity as a function of propagation distance under three different combinations of these two factors in Fig.\ref{figure4}.
It can be noticed in Fig.\ref{figure4} that the effect of crystal absorption on the THz intensity is more significant than the phase mismatch, i.e., the phase mismatch has almost no effect on the THz intensity when the propagation distance is not long enough. Since the absorption coefficient of PPKTP crystal in THz band is only 1.69 cm$^{-1}$ \cite{Kugel1988}, much lower than the values of its isomorphic crystals, the THz intensity in the PPKTP crystal is much higher than the value in  other crystals.

\begin{figure*}[tbh]
\centering\includegraphics[width=12cm]{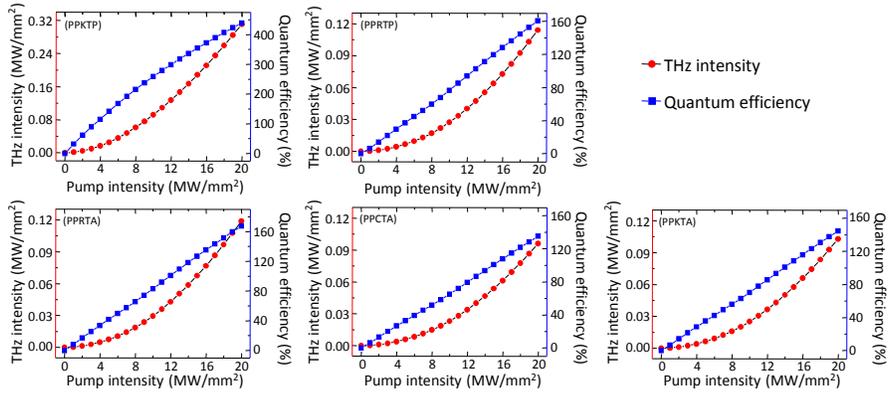}.
\caption{The maximum THz intensity and quantum conversion efficiency versus pump intensity for PPKTP and its isomorphic crystals (in the damage threshold of the crystals and considering 7$^{\rm{th}}$ order condition).
}\label{figure5}
\end{figure*}

In the processes of generating THz wave from periodically poled crystals, quantum conversion efficiency is a very important parameter. The higher the quantum conversion efficiency, the greater the THz intensity, and the more useful for practical applications.
Next, we simulate the quantum conversion efficiency in the cascaded DFG processes as a function of pump intensity in Fig.\ref{figure5} (considering 7$^{\rm{th}}$ order condition). The pump intensity is directly related to the THz intensity and quantum
conversion efficiency in the cascaded processes. As the cascade process continues, the pump intensity in the crystal gradually decreases. According to our and previous studies, the THz intensity will not change significantly after the 7$^{\rm{th}}$ cascade process. We change the intensity of the pump light from 1 MW/mm$^2$ to 20 MW/mm$^2$ with an interval of 1 MW/mm$^2$.

It can be noticed from Fig.\ref{figure5} that as the pump light intensity increases, the THz intensity
and the quantum conversion efficiency are also enhanced. When the pump light intensity reaches 20
MW/mm$^2$, the THz intensity and quantum conversion efficiency also reached their  maximum. According to
the calculation, the quantum conversion efficiency of PPKTP, PPRTP, PPKTA, PPRTA and PPCTA at 20 MW/mm$^2$
pump intensity can reach 438.5\%, 160.5\%, 144.7\%, 167.1\%, and 135.1\%, respectively, all exceed the
Manley-Rowe limit.
There are two reason reasons for this THz wave amplification during the DFG process: the first one is that
one pump photon can generate multiple THz photons in the cascaded DFG process, and the second reason is
that the three waves have a stronger interaction in the periodically poled crystal, due to the quasi-phase
matching technology.

\section{Discussion}
In Fig. \ref{figure3}, we only investigate the cases where the cascaded DFG is under the 9$^{\rm{th}}$-order.
For higher order cases, it can be calculated that the THz intensity does not increase any more. This is
mainly due to the following fact: when the propagation distance increases, the Stokes (anti-Stokes)
process decreased, at the same time, the crystal absorption of THz wave remains at a stable level.

The theoretical study in this article is based on room temperature. For periodically polarized crystals, temperature is a factor that cannot be ignored \cite{Jin2019JOLT, Emanueli2003, Yutsis2004, Mikami2011}.
The variation of temperature affects the refractive indices of the crystals, and thus changes the
quasi-phase matching conditions of the crystal. Besides, temperature changes also affect the absorption
coefficient of the pump light and the THz wave in the crystal, which are still in the stage of unexplored.
It is promising to select appropriate conditions by controlling the temperature to make the THz intensity
more efficient and stable.

KTP family have approximately 118 isomorphs \cite{Stucky1989, Sorokina2007, Gazhulina2013}, including 29
pure crystals and 89 doped crystals, all of which have a uniform form and can be written as $MM'OXO_4$,
where $\emph M$ = K, Rb , Na, Cs, TI, NH$_4$; $\emph M'$ = Ti, Sn, Sb, Zr, Ge, AI, Cr, Fe, V, Nb, Ta, Ga;
$\emph X$ = P, As, Si, Ge. Only five of these crystals are studied in this paper, and most of the other
crystals have not been thoroughly studied. Their Sellmeier equations, effective nonlinear coefficient and
absorption coefficient are still unknown. Therefore, exploring the nonlinear optical properties of these
crystals is necessary and may lead to more stable and stronger THz wave sources.

Here we discuss the limitation of this work. In the calculation, we mainly focus on the absorption effect of the different isomorphs in THz wave generation. However, some other effects also need to be taken into account. For example, the phase mismatch is affected by the dispersion of the pump in the medium. Furthermore, at high pump intensity, other limitations will occur, such as self-phase modulation (SPM) and cascaded effects. In the future, the theoretical model needs to be expanded to include these effects.

\section{Conclusion}
In conclusion, we have theoretically investigated the generation of THz wave from the isomorphs of PPKTP crystal, including PPRTP, PPKTA, PPRTA and PPCTA.  The process of generating THz wave by cascaded DFG is studied in detail, and the influences of different conditions on THz wave intensity are analyzed comprehensively. The results show that cascaded DFG method can effectively improve the intensity of THz wave; quantum conversion efficiency that exceeding the Manley-Rowe condition can be achieved at high pump intensity. In addition, we find that the impact of crystal absorption on THz wave intensity is more important than the phase mismatch. This study may provide more and stronger THz light sources based on the nonlinear optics crystals for future practical applications.

\section{Acknowledgments}
We thank Dr. Fabian Laudenbach for help full discussions. This work is  partially supported by the
National Natural Science Foundations of China (Grant Nos. 91836102, 11704290, 12074299, 61775025).

Conflict of interest statement:
The authors declared that they have no conflicts of interest to this work.
We declare that we do not have any commercial or associative interest that represents a conflict of interest in connection with the work submitted.

Supplementary Information:  Mathematica codes for the figures in the main text.


\begin{thebibliography}{99}
\newcommand{\enquote}[1]{``#1''}

\bibitem{Zhang2010book}
X.-C. Zhang and J.~Xu, \emph{Introduction to {THz} Wave Photonics} (Springer,
  2010), 2010th ed.

\bibitem{Lee2009}
Y.-S. Lee, \emph{Principles of {Terahertz} Science and Technology} (Springer,
  2009), 1st ed.

\bibitem{Zhong2011}
K.~Zhong, J.-Q. Yao, D.-G. Xu, H.-Y. Zhang, and P.~Wang, \enquote{Theoretical
  research on cascaded difference frequency generation of terahertz radiation
  (in {Chinese}),} {\protect\JournalTitle{Acta Phys. Sin.}} \textbf{60}, 034210
  (2011).

\bibitem{Kawase2007}
K.~Kawase and S.~Hayashi, \enquote{Terahertz wave parametric generation and
  applications,} {\protect\JournalTitle{Proc. of SPIE}} \textbf{6772},
  677202--(1--5) (2007).

\bibitem{Murate2018}
K.~Murate and K.~Kawase, \enquote{{Perspective: Terahertz wave parametric
  generator and its applications },} {\protect\JournalTitle{J. Appl. Phys.}}
  \textbf{124}, 160901 (2018).

\bibitem{Li2013}
Z.~Li, P.~Bing, D.~Xu, and J.~Yao, \enquote{High-power tunable terahertz
  generation from a surface-emitted {THz-wave } parametric oscillator based on
  two {MgO:LiNbO}$_3$ crystals,} {\protect\JournalTitle{Chin. Phys. B}}
  \textbf{124}, 4884--4886 (2013).

\bibitem{Burford2017}
N.~M. Burford and M.~O. El-Shenawee, \enquote{Review of terahertz
  photoconductive antenna technology,} {\protect\JournalTitle{Opt. Eng.}}
  \textbf{56}, 010901 (2017).

\bibitem{Du2015}
H.-W. Du, H.~Hoshina, and C.~Otani, \enquote{Thz generation from optical
  rectification tilted-pulse-front pumping scheme with laser pulse focused to a
  line,} {\protect\JournalTitle{Proc. SPIE}} \textbf{9671}, 96710M (2015).

\bibitem{Ding2014}
Y.~J. Ding, \enquote{Progress in terahertz sources based on
  difference-frequency generation,} {\protect\JournalTitle{J. Opt. Soc. Am. B}}
  \textbf{31}, 2696--2711 (2014).

\bibitem{Liu2013}
P.~Liu, D.~Xu, H.~Yu, H.~Zhang, Z.~Li, K.~Zhong, Y.~Wang, and J.~Yao,
  \enquote{Coupled-mode theory for cherenkov-type guided-wave terahertz
  generation via cascaded difference frequency generation,}
  {\protect\JournalTitle{J. Lightwave Technol.}} \textbf{31}, 2508--2514
  (2013).

\bibitem{Saito2015}
K.~Saito, T.~Tanabe, and Y.~Oyama, \enquote{Cascaded terahertz-wave generation
  efficiency in excess of the {Manley}-{Rowe} limit using a cavity
  phase-matched optical parametric oscillator,} {\protect\JournalTitle{J. Opt.
  Soc. Am. B}} \textbf{32}, 617--621 (2015).

\bibitem{Waldmueller2007}
I.~Waldmueller, M.~C. Wanke, and W.~W. Chow, \enquote{Circumventing the
  {Manley-Rowe} quantum efficiency limit in an optically pumped terahertz
  quantum-cascade amplifier,} {\protect\JournalTitle{Phys. Rev. Lett.}}
  \textbf{99}, 117401 (2007).

\bibitem{Hildenbrand2009}
A.~Hildenbrand, F.~R. Wagner, H.~Akhouayri, J.-Y. Natoli, M.~Commandr\'{e},
  F.~Th\'{e}odore, and H.~Albrecht, \enquote{Laser-induced damage investigation
  at 1064 nmin {KTiOPO}$_4$ crystals and its analogy with {RbTiOPO}$_4$,}
  {\protect\JournalTitle{Appl. Opt.}} \textbf{48}, 4263--4269 (2009).

\bibitem{Kim2015JKPS}
T.~Kim and K.~J. Lee, \enquote{Extended phase matching properties of
  periodically poled potassium titanyl phosphate isomorphs,}
  {\protect\JournalTitle{J. Korean Phys. Soc.}} \textbf{67}, 837--842 (2015).

\bibitem{Kim2021}
I.~Kim, D.~Lee, and K.~J. Lee, \enquote{Numerical investigation of high-purity
  polarization-entangled photon-pair generation in non-poled ktp isomorphs.}
  {\protect\JournalTitle{Appl. Sci.}} \textbf{11}, 565 (2021).

\bibitem{Jin2016PRAppl}
R.-B. Jin, P.~Zhao, P.~Deng, and Q.-L. Wu, \enquote{Spectrally pure states at
  telecommunications wavelengths from periodically poled {MTiOXO$_4$ (M = K,
  Rb, Cs; X = P, As)} crystals,} {\protect\JournalTitle{Phys. Rev. Appl.}}
  \textbf{6}, 064017 (2016).

\bibitem{Laudenbach2017}
F.~Laudenbach, R.-B. Jin, C.~Greganti, M.~Hentschel, P.~Walther, and H.~Hubel,
  \enquote{Numerical investigation of photon-pair generation in periodically
  poled { $M$TiO$X$O$_4$ ($M$=K, Rb, Cs; $X$=P, As)},}
  {\protect\JournalTitle{Phys. Rev. Appl.}} \textbf{8}, 024035 (2017).

\bibitem{Jin2019OLT}
R.-B. Jin, G.-Q. Chen, F.~Laudenbach, S.~Zhao, and P.-X. Lu, \enquote{Thermal
  effects of the quantum states generated from the isomorphs of {PPKTP}
  crystal,} {\protect\JournalTitle{Opt. Laser Technol.}} \textbf{109}, 222--226
  (2019).

\bibitem{McCracken2018}
R.~A. McCracken, F.~Graffitti, and A.~Fedrizzi, \enquote{Numerical
  investigation of mid-infrared single-photon generation,}
  {\protect\JournalTitle{J. Opt. Soc. Am. B}} \textbf{35}, C38--C48 (2018).

\bibitem{Li2017}
Z.-Y. Li, S.-L. Wang, M.-T. Wang, and W.-S. Wang, \enquote{Terahertz generation
  based on cascaded difference frequency generation with periodically-poled
  {KTiOPO}$_4$,} {\protect\JournalTitle{Curr. Opt. Photon.}} \textbf{1},
  138--142 (2017).

\bibitem{Li2017a}
Z.-Y. Li, M.-T. Wang, S.-L. Wang, D.-G. Xu, and J.-Q. Yao, \enquote{Highly
  efficient terahertz generation from periodically-poled {RbTiOPO}$_4$,}
  {\protect\JournalTitle{Optoelectron. Lett.}} \textbf{13}, 127--130 (2017).

\bibitem{Liu2013JLT}
P.~Liu, D.~Xu, H.~Yu, H.~Zhang, Z.~Li, K.~Zhong, Y.~Wang, and J.~Yao,
  \enquote{Theoretical analysis of terahertz generation in periodically
  inverted nonlinear crystals based on cascaded difference frequency generation
  process,} {\protect\JournalTitle{J. Lightwave Technol.}} \textbf{31},
  2508--2514 (2013).

\bibitem{Wang2015MRLB}
C.-F. Hu, K.~Zhong, J.-L. Mei, M.-R. Wang, S.-B. Guo, W.-Z. Xu, P.-X. Liu,
  D.-G. Xu, Y.-Y. Wang, and J.-Q. Yao, \enquote{Theoretical analysis of
  terahertz generation in periodically inverted nonlinear crystals based on
  cascaded difference frequency generation process,}
  {\protect\JournalTitle{Mod. Phys. Lett. B}} \textbf{29}, 1450263 (2015).

\bibitem{Lu2018OE}
W.~Lu, A.~Fallahi, K.~Ravi, and F.~K?rtner, \enquote{High efficiency terahertz
  generation in a multi-stage system,} {\protect\JournalTitle{Opt. Express}}
  \textbf{26}, 29744--29768 (2018).

\bibitem{Tian2021}
W.~Tian, G.~Cirmi, H.~T. Olgun, P.~Mutter, and etc, \enquote{$\mu${J}-level
  multi-cycle terahertz generation in a periodically poled {Rb}: {KTP}
  crystal,} {\protect\JournalTitle{Opt. Lett.}} \textbf{46}, 761--764 (2021).

\bibitem{Perkins1987}
P.~E. Perkins and T.~S. Fahlen, \enquote{20-{W} average-power {KTP}
  intracavity-doubled {Nd:YAG} laser,} {\protect\JournalTitle{J. Opt. Soc. Am.
  B}} \textbf{4}, 1066--1071 (1987).

\bibitem{Gettemy1988}
D.~J. Gettemy, W.~C. Harker, G.~Lindholm, and N.~P. Barnes, \enquote{Some
  optical properties of {KTP}, {LiIO}$_3$, and {LiNbO}$_3$,}
  {\protect\JournalTitle{IEEE J. Quantum Electron.}} \textbf{24}, 2231--2237
  (1988).

\bibitem{Kugel1988}
G.~E. Kugel, F.~Brehat, B.~Wyncke, M.~D. Fontana, G.~Marnier,
  C.~Carabatos-Nedelec, and J.~Mangin, \enquote{The vibrational spectrum of a
  {KTiOPO}$_4$ single crystal studied by {Raman} and infrared reflectivity
  spectroscopy,} {\protect\JournalTitle{J. Phys. C: Solid State Phys.}}
  \textbf{21}, 5565 (1988).

\bibitem{Shoji1997}
I.~Shoji, T.~Kondo, A.~Kitamoto, M.~Shirane, and R.~Ito, \enquote{Absolute
  scale of second-order nonlinear-optical coefficients,}
  {\protect\JournalTitle{J. Opt. Soc. Am. B}} \textbf{14}, 2268--2294 (1997).

\bibitem{Kato2002}
K.~Kato and E.~Takaoka, \enquote{Sellmeier and thermo-optic dispersion formulas
  for {KTP},} {\protect\JournalTitle{Appl. Opt.}} \textbf{41}, 5040--5044
  (2002).

\bibitem{Sang2010}
M.~Sang, J.-H. Qiu, T.-X. Yang, X.-C. Lu, and W.-L. Zhang, \enquote{Optical
  phonon resonance characteristics comparsion of {KTP} and {RTP} crystals in
  terahertz time-domain sperctroscopy,} {\protect\JournalTitle{Chin. J.
  Lasers}} \textbf{37} (2010).

\bibitem{Cheng1994}
L.~K. Cheng, L.~T. Cheng, J.~Galperin, P.~A.~M. Hotsenpiller, and J.~D.
  Bierlein, \enquote{Crystal growth and characterization of {KTiOPO}$_4$
  isomorphs from the self-fluxes,} {\protect\JournalTitle{J. Cryst. Growth}}
  \textbf{137}, 107--115 (1994).

\bibitem{Hansson2000}
G.~Hansson, H.~Karlsson, S.~Wang, and F.~Laurell, \enquote{Transmission
  measurements in {KTP} and isomorphic compounds,} {\protect\JournalTitle{Appl.
  Opt.}} \textbf{39}, 5058--5069 (2000).

\bibitem{Mikami2009}
T.~Mikami, T.~Okamoto, and K.~Kato, \enquote{Sellmeier and thermo-optic
  dispersion formulas for {RbTiOPO}$_4$,} {\protect\JournalTitle{Opt. Mater.}}
  \textbf{31}, 1628--1630 (2009).

\bibitem{Watson1991}
G.~H. Watson, \enquote{Polarized raman spectra of {KTiOAsO}$_4$ and isomorphic
  nonlinear-optical crystals,} {\protect\JournalTitle{J. Raman Spectrosc.}}
  \textbf{22}, 705--713 (1991).

\bibitem{Loiacono1992}
G.~M. Loiacono, D.~N. Loiacono, and J.~J. Zola, \enquote{Optical properties and
  ionic conductivity of {KTiOAsO}$_4$ crystals,}  (1992), p. CThD5.

\bibitem{Cheng1993a}
L.~K. Cheng, L.~T. Cheng, J.~D. Bierlein, F.~C. Zumsteg, and A.~A. Ballman,
  \enquote{Properties of doped and undoped crystals of single domain
  {KTiOAsO}$_4$,} {\protect\JournalTitle{Appl. Phys. Lett.}} \textbf{62},
  346--348 (1993).

\bibitem{Kato1997}
K.~Kato, N.~Umemura, and E.~Tanaka, \enquote{$90^{\circ}$ phase-matched
  mid-infrared parametric oscillation in undoped {KTiOAsO}$_4$,}
  {\protect\JournalTitle{Jpn. J. Appl. Phys.}} \textbf{36}, L403 (1997).

\bibitem{Mounaix2004}
P.~Mounaix, L.~Sarger, J.~Caumes, and E.~Freysz, \enquote{Characterization of
  non-linear {Potassium} crystals in the {Terahertz } frequency domain,}
  {\protect\JournalTitle{Opti. Commun.}} \textbf{242}, 631 -- 639 (2004).

\bibitem{Cheng1993b}
L.~T. Cheng, L.~K. Cheng, and J.~D. Bierlein, \enquote{Linear and nonlinear
  optical properties of the arsenate isomorphs of {KTP},}
  {\protect\JournalTitle{Proc. SPIE}} \textbf{1863}, 43--53 (1993).

\bibitem{Guo1996}
A.~R. Guo, C.~S. Tu, R.~Tao, R.~S. Katiyar, R.~Guo, and A.~S. Bhalla,
  \enquote{Temperature dependent raman scattering in {RbTiOAsO}$_4$ and
  {CsTiOAsO}$_4$ single crystals,} {\protect\JournalTitle{Ferroelectrics}}
  \textbf{188}, 143--156 (1996).

\bibitem{Yang1999}
Y.~Yang and C.~S. Yoon, \enquote{Dielectric properties of {RbTiOAsO}$_4$ single
  crystals,} {\protect\JournalTitle{Appl. Phys. Lett.}} \textbf{75}, 1164--1166
  (1999).

\bibitem{Kato2003}
K.~Kato, E.~Takaoka, and N.~Umemura, \enquote{Thermo-optic dispersion formula
  for {RbTiOAsO}$_4$,} {\protect\JournalTitle{Jpn. J. Appl. Phys.}}
  \textbf{42}, 6420 (2003).

\bibitem{Cheng1993}
L.~T. Cheng, L.~K. Cheng, J.~D. Bierlein, and F.~C. Zumsteg, \enquote{Nonlinear
  optical and electro-optical properties of single crystal {CsTiOAsO}$_4$,}
  {\protect\JournalTitle{Appl. Phys. Lett.}} \textbf{63}, 2618--2620 (1993).

\bibitem{Loiacono1993}
G.~M. Loiacono, D.~N. Loiacono, and R.~A. Stolzenberger, \enquote{Crystal
  growth and characterization of ferroelectric {CsTiOAsO}$_4$,}
  {\protect\JournalTitle{J. Cryst. Growth}} \textbf{131}, 323 -- 330 (1993).

\bibitem{Mikami2011}
T.~Mikami, T.~Okamoto, and K.~Kato, \enquote{Sellmeier and thermo-optic
  dispersion formulas for {CsTiOAsO}$_4$,} {\protect\JournalTitle{J. Appl.
  Phys.}} \textbf{109}, 023108 (2011).

\bibitem{Du2014}
H.-W. Du and N.~Yang, \enquote{Theoretical investigation on thz generation from
  optical rectification with tilted-pulse-front excitation,}
  {\protect\JournalTitle{Chin. Phys. Lett.}} \textbf{31}, 124201 (2014).

\bibitem{Cronin-Golomb2004}
M.~Cronin-Golomb, \enquote{Cascaded nonlinear difference-frequency generation
  of enhanced terahertz wave production,} {\protect\JournalTitle{Opt. Lett.}}
  \textbf{29}, 2046--2048 (2004).

\bibitem{Schaar2008}
J.~E. Schaar, K.~L. Vodopyanov, P.~S. Kuo, M.~M. Fejer, X.~Yu, A.~Lin, J.~S.
  Harris, D.~Bliss, C.~Lynch, V.~G. Kozlov, and W.~Hurlbut, \enquote{Terahertz
  sources based on intracavity parametric down-conversion in
  quasi-phase-matched gallium arsenide,} {\protect\JournalTitle{IEEE J. Sel.
  Top. Quantum Electron.}} \textbf{14}, 354--362 (2008).

\bibitem{Ravi2016}
K.~Ravi, M.~Hemmer, G.~Cirmi, F.~Reichert, D.~N. Schimpf, O.~D. M\"{u}cke, and
  F.~X. K\"{a}rtner, \enquote{Cascaded parametric amplification for highly
  efficient terahertz generation,} {\protect\JournalTitle{Opt. Lett.}}
  \textbf{41}, 3806--3809 (2016).

\bibitem{Jin2019JOLT}
R.-B. Jin, G.-Q. Chen, F.~Laudenbach, S.~Zhao, and P.-X. Lu, \enquote{Thermal
  effects of the quantum states generated from the isomorphs of {PPKTP}
  crystal,} {\protect\JournalTitle{J. Opt. Laser Technol.}} \textbf{109},
  222--226 (2019).

\bibitem{Emanueli2003}
S.~Emanueli and A.~Arie, \enquote{Temperature-dependent dispersion equations
  for {KTiOPO}$_4$ and {KTiOAsO}$_4$,} {\protect\JournalTitle{Appl. Opt.}}
  \textbf{42}, 6661--6665 (2003).

\bibitem{Yutsis2004}
I.~Yutsis, B.~Kirshner, and A.~Arie, \enquote{Temperature-dependent dispersion
  relations for {RbTiOPO}$_4$ and {RbTiOAsO}$_4$,} {\protect\JournalTitle{Appl.
  Phys. B}} \textbf{79}, 77--81 (2004).

\bibitem{Stucky1989}
G.~D. Stucky, M.~L.~F. Phillips, and T.~E. Gier, \enquote{The potassium titanyl
  phosphate structure field: a model for new nonlinear optical materials,}
  {\protect\JournalTitle{Chem. Mater.}} \textbf{1}, 492--509 (1989).

\bibitem{Sorokina2007}
N.~I. Sorokina and V.~I. Voronkova, \enquote{{Structure and properties of
  crystals in the potassium titanyl phosphate family: A review},}
  {\protect\JournalTitle{Cryst. Rep.}} \textbf{52}, 80--93 (2007).

\bibitem{Gazhulina2013}
A.~P. Gazhulina and M.~O. Marychev, \enquote{Pseudosymmetric features and
  nonlinear optical properties of potassium titanyl phosphate crystals,}
  {\protect\JournalTitle{Cryst. Struct. Theory Appl.}} \textbf{2}, 106--119
  (2013).

\end{thebibliography}
\end{document}